\documentstyle[12pt]{article}
\begin{document}
\vspace{.1in}
\def\oneskip{\vskip\baselineskip}
\def\bge{\begin{equation}}
\def\ede{\end{equation}}
\def\RR{\vbox {\hbox to 8.9pt {I\hskip-2.1pt ${\rm R}^3$\hfil}}}
\centerline{\large {\bf Force Error Optimization in Smooth Particle 
			Hydrodynamics}}

\vspace{.4in}

\centerline{$~^{1}$R. Capuzzo--Dolcetta, $~^{1,2}$R. Di Lisio }

\vspace{.1in}

\centerline{\footnotesize {\it $~^1$ Istituto Astronomico}}

\centerline{\footnotesize {\it Universit\`a di Roma ``La Sapienza"}}

\centerline{\footnotesize {\it Via G.M. Lancisi 29}}

\centerline{\footnotesize {\it I-00161  Roma}}

\vspace{.1in}

\centerline{\footnotesize {\it $~^2$ Dipartimento di Matematica}}

\centerline{\footnotesize {\it Universit\`a di Roma ``La Sapienza"}}

\centerline{\footnotesize {\it P.le Aldo Moro 5}}   

\centerline{\footnotesize {\it I-00185  Roma}}

\vfill



\eject

{\footnotesize
{\bf Abstract}
We discuss capability of Smooth Particle Hydrodynamics to represent
adequately the dynamics of self--gravitating systems, in particular for
what regards the quality of approximation of force fields in the 
motion equations. When cubic spline kernels are used, we find that
a good estimate of the pressure field cannot be obtained in 
non uniform situations using the commonly adopted scheme of
adapting the kernel sizes to include a fixed number of neighbours.
We find that a fixed number of neighbours gives the best approximation
of just the intensity of the force field, while the determination of the 
direction of the force requires a number of neighbours which strongly depends
on the particle position.
A good balance between quality of the approximation and
computer time consumption is obtained assuming
the latter about 2 times larger than the first one.
We suggest a procedure which is massively parallelizable.
With a suitable choice of the kernel sizes 
the amount of computations required is
less than twice the one required by the common SPH scheme at the
same level of approximation, so our recipe (when parallelized) 
is convenient.
}
\vspace{.4in}

\noindent
{\bf 1. Introduction}

\vspace{.2 in}

The subject of this paper is a numerical technique which is
largely used in fluid-dynamical
simulations: the so called Smooth Particle Hydrodynamics (SPH).

This particle method was firstly introduced by Lucy [1]
and Gingold and Monaghan [2] to simulate nonaxisymmetric fluids in
Astrophysics, and it became popular due to its robustness
and easy implementation.

Basically, the idea of the method is to consider
the fluid as an ensemble of (smooth) particles, anyone being representative
of a piece of fluid. In this scheme,
each particle is mathematically characterized by a ``kernel" (a symmetric, regular non-negative
function centered on the particle position)
which carries information on the average 
values of dynamical and thermodynamical quantities, as well as on their 
gradients. The size of the kernel may depend on the local density
of the particles.

Each particle moves in the force field generated by the whole particle
system, while the associated quantities evolve under their suitably 
regularized laws (a more detailed description of this method can be found
in [3]).

It is well known that necessary requirements for a good numerical method 
are compatibility with ``analytical" equations, convergence to the right 
solutions and stability.
At present,
these properties have not yet been completely proved for SPH. 
Few results are available; for instance Oelschl\"ager 
[4] and Di Lisio [5] have obtained
convergence results for a particular equation of state (EOS),
$P \propto \rho^2$,
in the free and self--gravitating cases, respectively.

A more general convergence result, for a  freely
moving polytropic compressible fluid,
can be found in Di Lisio, Grenier, Pulvirenti [6], [7]. 

These results seem to show that the SPH method has the desired characteristics
of convergence, at least for polytropic EOS. 

Of course, one of the most important requirement for the method to work well
is providing a good evaluation of the force field.

Usually, in the
simulations, the regularization kernels have compact support
(to reduce the computational
cost) and their widths are such that every particle interacts with
an almost constant and a priori fixed number of neighbours.
It is found that this evaluation strongly depends on the size of the kernels.

In this paper we attempt to give a better way to establish,
through a quantitative check of the error, a
correct size of the kernels for a good evaluation of the force field,
in the class of the cubic spline kernels. Actually, the spline kernels 
introduced and discussed by Monaghan and Lattanzio [8], are 
the most used in numerical simulations.

In what follows we shall denote with $\nabla$ or $\nabla_{\bf r}$
the usual simbolic vector of
components ${\partial\over \partial x},{\partial \over \partial y},
{\partial \over \partial z}$, while the symbol $*$ represent the usual
convolution operator in \RR.
The points in the space will be denoted
by ${\bf r}$ and $r\equiv |\bf r|$.
\oneskip\oneskip
\noindent
{\bf 2. SPH approximation of the fluid equations}
\oneskip
{\it 2.1 The free case}
\oneskip

A particle method to simulate the dynamical and thermodynamical evolution
of a fluid corresponds to a direct discretization of the Lagrangian set 
of fluid--dynamical equations. In the case, for instance, of a barotropic
compressible fluid without body forces this set writes as

\bge
\left \{ \matrix{ &{d \rho \over dt} + \rho \nabla \cdot {\bf v} =0 \cr
& {d {\bf v} \over dt} + {\nabla P \over \rho} =0 \cr
& P = P(\rho). }\right . 
\label{system}
\ede

\noindent
%where $\nabla$ is the usual simbolic vector of components ${\partial
%\over \partial x},{\partial \over \partial y},{\partial \over \partial z}$.
In the SPH scheme the fluid, whose total mass is $M$, is represented by an
ensemble of $N$ particles
with masses $m_i$ ($i=1,\dots,N$) such that $\sum_i m_i=M$, centered at points
$\bf r_i$; $\{ \bf r_i  \}$ is a set of space points sampling 
the volume occupied by the fluid.
To each particle is associated a density function called ``kernel". The most
commonly used kernel is the third order spline function (see [8])
\bge
\phi({\bf r},h)={1 \over  \pi h^3  }
\left \{ \matrix {& 1 - ({{3v^2} /2}) + ({3v^3 / 4}) &if~~0 
\leq v \leq 1\cr
& ( 2 - v )^3/4 &if~ 1 \leq v \leq 2 \cr
& 0 & elsewhere} \right .
\label{kernel}
\ede
where $v=r/h$. The width of the kernel, $2h$, is the radius of the
spherical ``fluid particle''; this radius may change particle by particle.

In SPH the pressure field per unit mass in the Euler equation is
usually approximated by

\bge
\left ( {\nabla P \over \rho}
\right )_{SPH}({\bf r}_i)=
\sum_{j=1}^N \left( {P(\rho_i)\over \rho_i^2} + {P(\rho_j)\over \rho_j^2}
\right ) \nabla \phi({\bf r}_i-{\bf r}_j,h_i),
\label{pressure1}
\ede
\noindent
where

\bge
\rho_i\equiv \rho({\bf r}_i)= \sum_{j=1}^N m_j \phi({\bf r}_i-{\bf r}_j,h_i)
\label{pressure2}
\ede
\noindent
is the SPH approximation to the fluid mass density.

The density (\ref{pressure2})
evolves in time because the particles move under the Newton law
\bge
{\ddot {\bf r}}_i = - \left ( {\nabla P \over \rho} \right )_{SPH}({\bf r}_i)
~~~~~~~~i=1,\cdots,N
\ede
which is the SPH version of the momentum equation in
the system (\ref{system}).

\oneskip
\noindent
{\it 2.2 The self--gravitating case}
\oneskip
SPH methods developed mainly in Astrophysics, due to its ability to handle
fully 3D situations. Most of astrophysics simulations should necessarily
include self-gravity. The system of equations underlying the motion of
barotropic self-gravitating fluids is a little more complicated than (1):
\bge
\left \{ \matrix{ &{d \rho \over dt} + \rho \nabla \cdot {\bf v} =0 \cr
& {d {\bf v} \over dt} + {\nabla P \over \rho} - \nabla U=0 \cr
& \nabla \cdot \nabla U = -4 \pi G \rho \cr
& P = P(\rho) }\right .
\ede
where $G$ is the gravity constant. Due to the spherical simmetry
of the kernels and Newton theorems, the SPH approximation of the
gravitational force field is simply:
\bge
\left ( \nabla U \right )_{SPH}
( {\bf r}_i ) = -\sum_{j=1}^N 4 \pi G{({\bf r}_i- {\bf r}_j)
\over | {\bf r}_i - {\bf 
r}_j |^3} \int_0^{|{\bf r}_i - {\bf r}_j|} \phi(x,h_i) x^2 dx
\label{gravsum}
\ede
This SPH approximation is quite accurate, and numerical simulations
show that the
gravitational field is always fitted better than the pressure one (because
of that the sum in (\ref{gravsum}) is usually substituted by a further
approximation as in the P3MSPH code suggested by Monaghan and Lattanzio [8]
or in TREESPH code proposed by Hernquist and Katz [9]).
It is easy to convince oneself about that when considering the different nature
of the pressure and gravitational fields. The first depends on the local
distribution of the matter, and so it is strongly subject to the local
fluctuations of the particle distribution, while the second rather
depends on the global matter density. This is the reason why, in a 
particle scheme, it is easier
to fit the gravitational force field than the pressure one.

So, in this paper we will focalize our attention on the SPH
approximation of the pressure field, only.

\oneskip
\noindent
{\it 2.3 Evaluating the error}
\oneskip

The $h$ value determines the number of
particles which give an effective contribution in the sums $(\ref{pressure1})$
and $(\ref{pressure2})$,
that is the
number of neighbouring particles that effectively interact (touch) with the
given one.
This means that the effective pressure force strongly depends on $h$:
a small $h$ means a steep  $\phi$ profile in (\ref{kernel})
and a small number of
particle contributing to the sums, and viceversa for large $h$.
It is easy to verify the two limits:
\bge
\lim_{h \to 0} \left({ {\nabla P \over \rho}}\right)_{SPH}({\bf r}) =0~~~~~~~
\lim_{h \to \infty} \left({ {\nabla P \over \rho} }\right)_{SPH}({\bf r}) =0.
\label{limits}
\ede
In the common case of a polytropic EOS ($P\propto \rho^\gamma$)
the second limit holds just for $\gamma > 2/3$.
It is so quite clear how the kernel size $h$ is crucial to give a good
SPH approximation of the pressure force field so that the particles,
fluid--representatives, move on the correct trajectories (which are the
characteristics of the fluid).

A quite natural way to determine a ``best" local value of $h$ is through
the minimization of the absolute value of the relative error 
\bge
E_r({\bf r};h)= {\left|{ 
\left({ {\nabla P \over \rho} }\right) ({\bf r}) - 
\left({ {\nabla P \over \rho} }\right)_{SPH}  ({\bf r}) 
}\right|
\over 
\left|{ \left({ {\nabla P \over \rho} }\right) ({\bf r})
}\right|}
\label{relerror}
\ede
At a given particle configuration, for fixed $\bf r$,
$E_r$ depends on $h$ only.

If we set the unitary vectors ${\bf n}_e({\bf r})=
({ {\nabla P \over \rho} })/
|{ {\nabla P \over \rho} }|$ (the exact direction) and 
${\bf n}_a({\bf r})=({ {\nabla P \over \rho} })_{SPH}/
|{ {\nabla P \over \rho} }|_{SPH}$ (the approximated SPH direction)
the relative error (\ref{relerror}) can be rewritten as
\bge
E_r({\bf r};h) =\left| ({\bf n}_e - {\bf n}_a) +
{|{ {\nabla P \over \rho} }|-
|{ {\nabla P \over \rho} }|_{SPH}
\over |{ {\nabla P \over \rho} }|}
{\bf n}_a \right|
\label{globerror}
\ede
This obvious identity suggests us to split the relative error into a
{\it direction} error $E_{dir}$ and a {\it modulus} error $E_{mod}$.
Respectively
\bge
E_{dir}({\bf r};h)= |{\bf n}_e - {\bf n}_a|~,~
E_{mod}({\bf r};h)= \left | 
{|{ {\nabla P \over \rho} }|-
|{ {\nabla P \over \rho} }|_{SPH}
\over |{ {\nabla P \over \rho} }|}
{\bf n}_a \right |
\label{errdefinitions}
\ede
\oneskip\oneskip
\noindent
{\bf 3. A model for comparing exact and SPH pressure fields}
\oneskip

To test the sensitivity of the SPH approximation of the force field on
the kernel width, we will consider particle distributions sampled from the
family of
Plummer's density laws [10]:
\bge
\rho(r)={\rho_0 \over \left [1 + \left ( {r / r_c} \right )^2 \right ]
^{5/ 2}}. 
\label{denmodel}
\ede
The law (\ref{denmodel}) is often used in theoretical Astrophysics because
it is a good 2-parameter ($\rho_0$ and $r_c$) fit to many self-gravitating
objects (globular clusters, roundish elliptical galaxies, etc.).
Here it is used just as a
reference density law apt to simulate a range of different density and
pressure gradients. To do this one can change the central density $\rho_0$
and the lenght scale $r_c$; increasing $r_c$ means dilute the distribution
(smoothing the gradients).

We fix the total mass of the sphere, $M$ so that the central density
$\rho_0$ is linked to the characteristic length scale $r_c$ which is
our second free parameter.
As we said, in the SPH scheme the density is represented by an ensemble of 
$N$ particles
with masses $m_i$ and positions ${\bf r}_i~(i=1,\dots,N)$.
In this scheme the pressure field in the ${\bf r}_i$ position can be 
approximated by (\ref{pressure1}).
The particle distribution sampling the regular density (\ref{denmodel})
is obtained
choosing first a set of random particle positions
${\overline {\bf r}}_i$ according to a 
uniform probability law in a sphere and then replacing every ${\overline{\bf 
r}}_i$ by ${\bf r}_i$ such that
\bge
\int_0^{r_i} \rho(s) s^2 ds = {{\overline r}_i^3 \over 3}.
\ede
%In the SPH scheme, to each particle is associated a mass $m_i$ and a
%kernel
%$$
%\phi(r,h)={1 \over  \pi h^3  }
%\left \{ \matrix {& 1 - ({{3v^2} /2}) + ({3v^3 / 4}) &if~~0 
%\leq v \leq 1\crIt is found that this evaluation strongly depends on the size of the kernels. 
%& ( 2 - v )^2/4 &if~ 1 \leq v \leq 2 \cr
%& 0 & elsewhere} \right .
%\eqno (2)
%$$
%where $v=r/h$.

%The pressure force field per unit mass is defined by
%$\vec F = \vec \nabla P/ \rho$.

In our simulations we shall use $10,000$ particles of individual
mass $m_i=M/10,000$.
%The constant $h_i$ is assumed to be the same for all the particles.

An interesting quantity associated to $h$ is the number of
particles which give an effective contribution in the sums $(\ref{pressure1})$
and $(\ref{pressure2})$, that is the
number of particles that effectively interact with the given one. In practical
simulations, this number is usually set to a constant value, the same for every
particle.

Due to the spherical simmetry of our model, the directional part of the
relative error $E_{dir}$ is a decreasing function of $h$,
and is obviously minimized 
taking $h$ as large as possible (which is computationally expensive),
while the error in the absolute value $E_{mod}$ is not a monotonic function
of $h$ (see eq.(\ref{limits})).
So, it is quite natural to study the modulus part of the 
error separately.

\oneskip\oneskip

\noindent
{\it 3.1 The results}
\oneskip

We have studied the minimization of the relative error $E_r$, as a function
of $h$ (assumed to be the same for each particle)
for three different values of $r_c$ in (\ref{denmodel}), namely
$0.25$, $0.5$, $1$.
For $\gamma$ the values $7/5$, $5/3$ and $2$, of clear 
physical interest, are considered. 

For each case the minimization of $E_r$ as function of $h$
has been performed on a set of 100 different particle positions.
To do that we evaluate the pressure field on each particle using the same
kernel for all the particles, then we modify the size of this kernel
to obtain the best result.

Fig.s 1 and 2 show (for some $r_c$ and $\gamma$) the minimum relative
error $E_r$ as function of the distance from the centre
and the number of particles (neighbours) within the kernel of any of
the 100 particles, respectively.

These figures seem to show that the best results (low value for $E_r$
coupled with a small number of neighbours) have been obtained in the
middle radius zone.
The minimum error $E_r$ is, in the average, a slightly decreasing function
of $r$.

To study better the behaviour of the $E_r$ function, we have done
a separate statistics of
the modulus ($E_{mod}$) and directional ($E_{dir}$) errors.
We have considered
30 particles in three different regions, around $r/r_c=0,~1,~2$,
assuming, in the state law, $\gamma = 5/3$.
To give a visual idea of the quality of the SPH approximation,
in Fig.s $3,~4,~5$ the SPH pressure field is compared with the exact one. 
For graphical convenience, for each particle
we have plotted the exact vector and a vector whose modulus is given by the best
SPH approximation (that given by $E_r$ minimization) at an angle, clockwise
oriented, given by the scalar product between the exact, ${\bf n}_e$,
and SPH, ${\bf n}_a$, unitary vectors in space (cp. Sect. 2.3).
The exact vector points to the centre; in the figures the centre
is assumed to be at an infinite distance.
It is quite evident the better field direction approximation for larger
$r_c$ and $r/r_c$ (i.e. less steep gradients).

To give a more precise measure of the errors, in Fig.s $6,~7,~8$
we have plotted the relative error $E_r$ and the
modulus error $E_{mod}$. The number of neighbours and,
to give a measure of the
direction error $E_{dir}$, the cosine of the angle $\Theta$
between the SPH and exact vector fields
($\cos\Theta = {\bf n}_e\cdot {\bf n}_a$) have also been plotted.
Fig.s $6-8$ show
that the field is well approximated in the middle zone while 
a very good approximation of just the direction of the
pressure field is obtained in the outer one.
The best balancing of the two errors have been obtained in the middle zone.
In the inner zone, where the pressure gradient should almost vanishing, the
approximation is critically dependent on sampling problems (see discussion
in Sect. 4.).

The different radial behaviours of the errors $E_r$, $E_{mod}$ and
$E_{dir}$, suggest us to split the {\bf evaluation} of SPH approximation
in two parts: 

\noindent
first we fix $h$ to evaluate at best the modulus of the pressure field, then
we enlarge it to obtain an acceptable direction of the field
with a computationally reasonable number of neighbours.

%To approach this point a theoretical discussion is needed. Formula (3) can
%be seen as a weighted average of the directions ${\bf n}_{i,j}=
%({\bf r}_i-{\bf r}_j)/|{\bf r}_i-{\bf r}_j|$. The weights are
%$$
%\left( {P(\rho_i)\over \rho_i^2} + {P(\rho_j)\over \rho_j^2}
%\right ) \left . {d\phi \over dx}\right|_{|{\bf r}_i-{\bf r}_j|}
%|{\bf r}_i-{\bf r}_j|.
%$$
%The sum of the weights should be 1. But this not verified, for example,
%in the linear case (i.e. $P \propto \rho^2$). In this last case a condition
%on the kernel have to been assumed, that is $\int \phi'(x) |x| =1$.
%The cubic spline kernels here used are not of this kind.

To this scope, for several particles positioned at different radii, we have
plotted the errors $E_r$, $E_{mod}$ and $E_{dir}$ as functions of $h$.
These plots, Fig.$9$, show that
the best error in the pressure field absolute value ($E_{mod}$) is reached
with a smaller $h$ value than when the global error ($E_r$) was minimized.

Taking into account these results, we suggest to estimate separately the 
modulus and the direction
of the pressure field using two different kernel widths,
as aforementioned. The first width, say $h_{mod}$, is obtained
through the minimization of the modulus error $E_{mod}$,
the second, $h_{dir}$, is taken
proportional to the first in a way to keep the error in direction $E_{dir}$
below an acceptable limit.

To set such proportionality constant, we evaluate, for any particle
in the three regions considered before, the quantity $E_{dir}$
as function of the variable $h/h_{opt}$, where $h_{opt}$ is the width
obtained minimizing the modulus error $E_{mod}$.

The plots in Fig. $10$ have been obtained
averaging the $E_{dir}(h/h_{opt})$ functions found for each particle.
The lenght of the vertical
straight lines is proportional to the dispersion around the
mean value. 

This figure shows that it is sufficient to double the $h_{opt}$
value to get a good (within $10 \%$)
approximation of the direction of the field. So a reasonable
choice of the proportionality constant is $2$.

In Fig.s $11,~12,~13$ comparisons of the results obtained following
this recipe 
compared with those obtained by the simple minimization of $E_r$ are shown.
In these figures the number of neighbours corresponding to
$h_{opt}$ and the percentage global error $E_r$ are plotted for all
particles in the three zones before considered for $r_c=0.25,~0.5,~1$,
respectively, and $\gamma=5/3$.
The figures show that $h_{opt}$ is associated to
an almost constant number of neighbour particles, which corresponds 
to about $10 \%$ of the
total number. This behaviour occurs in all the cases here considered.

Note that usual SPH implementations evaluate the pressure field using
kernels with a size such that the number of neighbouring particles is
set to a given constant.
Thus, we tested the computational
cost of the implementation here proposed by comparing the CPU times
(on a DEC Alpha 200 4/233) required to evaluate the pressure field
by the usual and our SPH method.

We have fixed the number of neighbouring particles
to compute the absolute value of the pressure gradient to $10 \%$ of the
total, as suggested by our $E_{mod}$ minimization. This fraction corresponds
to a certain kernel size, and to evaluate the direction of the field we double
this size. In the usual method just one kernel size is used (here we
choose the one giving $10\%$ neighbours).

As axpected, the computational cost of a double-kernel evaluation of the
pressure field is $\approx 1+2^3$ times the single-kernel one, which is
the price of the good improvement of the error.

For the sake of a good balance between computational weight and precision
we tried $h_{dir}=\root 3 \of 2 h_{mod}$, i.e. the size needed to almost double the
number of neighbours. This case requires a computational cost which is
of the order of $1+2$ times the single-kernel method with $h=h_{mod}$, with
a lower error level; to have a similar approximation we verified that
$h_{mod}$ is allowed to be $<h$ thus to give a ratio of CPU times less than
2. Being this part of SPH evaluation wholly parallelizable explains why
in this way we can obtain a better approximation waiting the same solar time.
\oneskip\oneskip
\noindent
{\bf 4. Comparing SPH and Monte-Carlo methods}
\oneskip

The basic idea of the SPH method is to ``reconstruct" a function $f$,
whose values are known in some ``particle" points, by interpolation
(in fluid-dynamical
applications the particles are distributed to fit the fluid mass density).
The way SPH
does that is through a Monte-Carlo approximation of the convolution
integral $f*\phi$,
where $\phi$ is a regular
kernel function.
The function $f*\phi$ is close
to $f$
if $\phi$ is close to the Dirac's delta. In the same way it is possible 
to reconstruct $\nabla f$. Indeed, by noting that $(\nabla f) * \phi =
f * (\nabla \phi)$, we can evaluate the latter integral, instead of
the first, through the knowledge of $f$ only.
A similar procedure may be used for any derivative of $f$ with respect to
space variables.

Monte-Carlo methods are the most studied and the most advanced
particle methods in numerical analysis, so it is quite natural to compare
the SPH approximation method with direct Monte-Carlo ones.

While we refer the reader to a forthcoming paper [11] for a more detailed
and accurate analysis of the link between SPH and Monte-Carlo approximation
methods, for the purposes of this paper we limit here ourselves to show the
difference between the SPH
and the Monte-Carlo interpolations for the pressure force field only.

As we said in previous Sections,
the pressure force field $\nabla P/\rho$ is the most delicate term to handle in
fluid-dynamical simulations. In SPH, $\nabla P/\rho$ is usually approximated
by eq.(\ref{pressure1}).
The Monte-Carlo rule, choosing $\rho$ as distribution function
of the particles, gives
\bge
\left ({\nabla P \over \rho}\right )_{MC} ({\bf r})=
{1\over N} \sum_i {\nabla P({\bf r}_i) \over \rho^2({\bf r}_i)}
\phi ({\bf r}-{\bf r}_i,h),
\label{mc}
\ede
where we assume $h_i=h$ for any $i$.
The average value, over the particle configuration space, of the sum
in the right hand size of eq.(\ref{mc}) gives
the exact pressure field $(\nabla P
/ \rho)({\bf r})$. On the contrary, averaging the SPH approximation sum
(\ref{pressure1}) we find 
\bge
\left \langle \left ( {\nabla P\over \rho} \right )_{SPH} \right \rangle 
({\bf r})
= \left ( {\nabla P\over \rho} \right ) ({\bf r})+
\int \left ( {P({\bf r})\over \rho^2({\bf r})} -
{P({\bf r'})  \over \rho^2({\bf r'})}\right ) \nabla \rho({\bf r'})
\phi({\bf r}-{\bf r'},h) d^3{\bf r'},
\ede
the value of the integral being zero if $P \propto \rho^2$, otherwise it
goes to zero with $h$ whenever the functions
$\nabla P / \rho^2$ and $\nabla \rho$ are regular enough (for example
if they are differentiable). This means
that the SPH scheme (without artificial viscosity) may not be apt to
represent extreme situations, as when,
for example, shock fronts occur (i.e. when a discontinuity on the density
occurs).

Now, it is easy to show that the difference $\Delta$ between the 
Monte-Carlo and SPH sums writes as
$$
\Delta=
{1\over N} \sum_i \left [ \nabla_{\bf r'} \left ( {P({\bf r'}) \over
\rho^2({\bf r'})}
\phi({\bf r}-{\bf r'},h) \right ) \right ]_{{\bf r'}={\bf r}_i} + 
$$
\bge
{2 \over N} \sum_i
{P({\bf r}_i) \over \rho^3({\bf r}_i)} \nabla \rho({\bf r}_i)
\phi({\bf r}-{\bf r}_i,h)
-{1\over N} \sum_i {P({\bf r}) \over \rho^2({\bf r})} \nabla_{\bf r}
\phi ({\bf r}-{\bf r}_i,h)
\label{diff1}
\ede
and its average value is not zero, as we said.

This term is too difficult to handle in general situations. But we can study
it in some simple cases. For example, for a fluid which state law is
$P\propto \rho^2$, then eq.(\ref{diff1}) reduces to
\bge
\Delta=
{2 \over N} \sum_i {1 \over \rho({\bf r}_i)} \nabla_{{\bf r}_i}
\left (\rho({\bf r}_i) \phi({\bf r}-{\bf r}_i) \right ).
\label{simple1}
\ede
In a situation not far from spatial uniformity
($|\nabla \rho |/\rho << 1$),
eq.(\ref{simple1}) reduces to:
\bge
\Delta\simeq {2 \over N} \sum_i  \nabla_{{\bf r}_i} \phi({\bf r}-{\bf r}_i),
\ede
whose averaged value is the vector $-\nabla \rho$.
On the other hand, if we suppose  $|\nabla \rho |/\rho >> 1$
\bge
\Delta
\simeq
{2 \over N} \sum_i {\nabla_{{\bf r}_i}
\rho({\bf r}_i) \over \rho({\bf r}_i)}  \phi({\bf r}-{\bf r}_i)
\ede
holds, and the averaged value of the right hand term
is the vector $\nabla \rho$.

The first case should be the better one, anyway when the density
profile is quasi-constant, the
fluctuations due to the particle fit amplify the relative error.

Let us consider now the state law $P \propto \rho^{2+\alpha}$ where
$\alpha > 0$. When
$\ln^2(\rho) max(1, \rho^{-\alpha}) << 1$ we have
$P\simeq \rho^2 (1+ \alpha \ln(\rho))$, and (\ref{diff1}) writes as
$$
\Delta \simeq {2 \over N} \sum_i {1 \over \rho({\bf r}_i)} \nabla_{{\bf r}_i}
\left (\rho({\bf r}_i) \phi({\bf r}-{\bf r}_i) \right )
+
$$
\bge
{2 \alpha \over N} \sum_i \nabla_{{\bf r}_i}
\left [\ln(\rho({\bf r}_i)) \phi({\bf r}-{\bf r}_i) \right ]
+
{ \alpha \over N} \sum_i \phi({\bf r}-{\bf r}_i)
\nabla_{{\bf r}_i} \ln^2(\rho({\bf r}_i)).
\ede
Let us study again the two 
extreme cases $|\nabla \rho|/\rho
<< 1$ and $|\nabla \rho|/\rho>>1$. It is easy to convince ourselves that
the previous considerations still hold, but only if $|\ln \rho|$
is not too great. This means, for example, that SPH
approximation is not expected to be accurate near the boundaries of the fluid.

In conclusion,
the best SPH fit (i.e. comparable with Monte-Carlo one) for the pressure
field can be obtained when both
the following conditions hold: 

i) the density is not too large or too small
and ii) it has a profile along a well defined direction with small gradients.

Moreover, we notice that the previous analysis does not depend on the
overall mass distribution
but only on its local behaviour. Then the conditions i) and ii) are valid
criteria in very general situations.

Let us now apply previous considerations to the spherical
model studied in this paper.
The results
of our simulations, shown in Sect. 3, confirm the expectations based on
the previous analisys:
the best results on the evaluation of the pressure
field are obtained in those zones of the model where the conditions i)
and ii) approximately hold, i.e. in the central ones as suggested by
the density law (\ref{denmodel}) and
\bge
{|\nabla \rho| \over \rho} \propto {r\over(r^2+r_c^2)}
\ede
Moreover, comparing models with different
characteristic radii we found that the precision increases as $r_c$ grows,
as suggested by 
\bge
\max_r {|\nabla \rho| \over \rho} \propto {1 \over r_c}.
\label{logder}
\ede

\oneskip\oneskip
\noindent
{\bf 5. Conclusions}
\vspace{.2 in}

In this paper we have studied the error in the SPH evaluation
of the pressure field
for the set of polytropic compressible fluids in spherical simmetry.

\noindent
The spherical simmetry allows us to compare
SPH evaluation with a simple model, saving the confidence
that the main results are still valid in more general and realistic 
situations, as we have shown with analytical considerations in Sect.4.
First we determined the kernel sizes which minimize the global error
$E_r$ (see eq.(\ref{relerror}));
they depend on the particle positions becoming
narrower where the density is lower, due to that in external regions the
right (radial) direction of the field is easily obtained.
This latter is linked to that a good approximation of the pressure field is
harder to be obtained in very central and peripheral regions. Actually,
in very central zones, where $|\nabla P| / \rho \simeq 0$ and $|\nabla \rho|
/ \rho<< 1$, to get an acceptable fit of the pressure field direction so
large kernel sizes are nedeed to make the SPH pressure gradient
underestimated. On the contrary, in the outer regions, the radial
direction of the field is well fitted even with small kernel sizes
(i.e. few neighbours): this implies a certain overestimate of the
pressure gradient.

All this has suggested us to study separately the behaviours of the modulus
and directional error, as defined by eq. (\ref{errdefinitions}).

\noindent
Given that the larger the kernel the better the approximated direction,
we minimized the error in modulus $E_{mod}$
rather than the global one.

The obtained kernel width slightly depends on the particle position thus
an almost constant number of neighbours is needed to perform the best
approximation. 
In our simulations this number is found to be $10 \%$ of the total
independently of the particle positions and the characteristic
radius of the model.

Moreover, a good (giving an error less
than $10 \% $) direction of the pressure field
is obtained doubling that kernel size.
Unfortunately, this evaluation requires a CPU time almost 8 times the
one required by the
single-kernel evaluation. 
We have obtained a good balance between the 
quality of approximation and the CPU weight
fixing the `direction' kernel to be $\root 3 \of 2$ times the
`modulus' one.

So, a good prescription to estimate the pressure field in SPH is
to evaluate the absolute value of the field with a kernel size giving
a constant number of neighbouring particles, while the direction
is estimated using another kernel size as described before.

The CPU time required by this
procedure is less than two times that required by the usual
(single-kernel) SPH
evaluation of the pressure field. This recipe becomes convenient when
a parallelized code is used.

\vfill\eject

\centerline{\bf References}

\vspace{.2 in}

{\footnotesize

\begin{enumerate}

\item Lucy, L., {\it Astron. J.}, {\bf 82}, 1013 (1977).
\item Gingold, R.A., Monaghan, J.J., {\it Mon. Not. Roy. Astron. Soc.},
{\bf 181}, 375 (1977).
\item Monaghan, J.J., {\it Ann. Rev. of Astron. and Astroph.},
{\bf 30}, 543 (1992).
\item Oelschl$\ddot {\rm a}$ger, K., {\it Arch. Rat. Mech. An.}, {\bf 115},
297 (1991).
\item Di Lisio, R., {\it Mat. Met. Appl. Sci.}, {\bf 18}, 1083 (1995).
\item Di Lisio, R., Grenier, E. and Pulvirenti, M., {\it Special Issue
devoted to Simulations Methods in Kinetic Theory}, Computers and
Mathematics with Applications, in press (1996).
\item Di Lisio, R., Grenier, E. and Pulvirenti, M.,
{\it Annali della Scuola Normale di Pisa}, in press (1996).
\item Monhagan, J.J., Lattanzio, J.C., {\it Astron. Astrophys.}, {\bf 149},
135 (1985).
\item Hernquist, L., Katz, N., {\it Ap. J. Suppl.}, {\bf 70}, 419 (1989).
\item Plummer, H.C., {\it Mon. Not. Roy. Astron. Soc.}, {\bf 71}, 460 (1911).
\item Capuzzo-Dolcetta, R., and Di Lisio, R., in preparation.

\end{enumerate}
}
\vfill \eject

\centerline{\bf Figure captions}

\vspace{.3in}

\centerline{\bf Fig. 1}

{\footnotesize
\noindent

For the Plummer's sphere with $r_c=0.25$, and polytropic exponent
$\gamma = 7/5$ (solid line) and $\gamma=2$ (dashed line)
panel (a) shows the minimum relative error $E_r$ (in percent) vs.
the radial coordinate and panel (b) the corresponding relative percentage
number of neighbouring particles $N_{n}$ (see text). 
}

\vspace{.2in}

\centerline{\bf Fig. 2}

{\footnotesize
\centerline{
As in Fig.1 but $r_c=1$.}
}

\vspace{.2in}

\centerline{\bf Fig. 3}

{\footnotesize
\noindent

For the polytropic exponent $\gamma=5/3$, the exact (vertical arrows)
and approximate $\nabla P/\rho$ fields are shown in three different
radial zones, for $r_c=0.25$ .
}

\vspace{.2in}

\centerline{\bf Fig. 4}

{\footnotesize
\centerline{
As in Fig.3 but for $r_c=0.5$}
}

\vspace{.2in}

\centerline{\bf Fig. 5}

{\footnotesize
\centerline{
As in Fig.3 but for $r_c=1.$}
}

\vspace{.2in}

\centerline{\bf Fig. 6}

{\footnotesize
\noindent

For the polytropic exponent $\gamma=5/3$ and $r_c=0.25$,
we show, from bottom up:

- the minimum relative error $E_r$ in percentage;

- the corresponding modulus error $E_{mod}$ in percentage;

- the cosine of the angle $\Theta$ between the unitary vectors
${\bf n}_a$ and ${\bf n}_e$ (see text);

- the number of neighbours $N_{n}$ in percentage to the total;

\noindent
for all the 30 particles in each of the three chosen shells.
Solid line refers to the
inner ($r/r_c=0$) shell, dashed line to the middle ($r/r_c=1$) and
dot-dashed to the outer ($r/r_c=2$).
}

\vspace{.2in}

\centerline{\bf Fig. 7}

{\footnotesize
\centerline{
As in Fig.6 but for $r_c=0.5$}
}

\vspace{.2in}

\centerline{\bf Fig. 8}

{\footnotesize
\centerline{
As in Fig.6 but for $r_c=1.$}
}

\vfill\eject

\centerline{\bf Fig. 9}

{\footnotesize
\noindent
For the polytropic exponent $\gamma = 5/3$ and characteristic radius
$r_c=0.5$, the values of the errors $E_r$ (solid lines),
$E_{mod}$ (dashed lines) and
$E_{dir}$ (dot-dashed lines) are plotted as functions of $h/r_c$ for
10 particles positioned at different distances from the centre (as
labelled).
}

\vspace{.2in}

\centerline{\bf Fig. 10}

{\footnotesize
\noindent

For the polytropic exponent $\gamma=5/3$ and $r_c=0.25,~ 0.5,~ 1$ the
averaged (over all the 30 particles in each shell) directional error $E_{dir}$
as function of $h/h_{opt}$ is plotted (see text). The lenghts of the
vertical straight lines are proportional to the dispersions of the
data around the mean values.
}

\vspace{.2in}

\centerline{\bf Fig. 11}

{\footnotesize
\noindent

For the polytropic exponent $\gamma=5/3$ and
$r_c=0.25$ the percentage
number of neighbours $N_{n}$ and the percentage total error $E_r$
are shown for all the particles in the shells when our double-kernel
method is applied
(solid lines) and when $E_r$ is minimized (dashed lines) for
all the 30 particles in the shells (as lebelled in the right panels). In
each case the horizontal straight line correspond to the average
value over all the 30 particles.

}

\vspace{.2in}

\centerline{\bf Fig. 12}

{\footnotesize
\centerline{
As in Fig.11 but for $r_c=0.5$}

}

\vspace{.2in}

\centerline{\bf Fig. 13}

{\footnotesize
\centerline{
As Fig.11 but for $r_c=1.$}

}
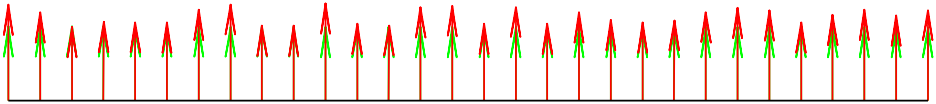
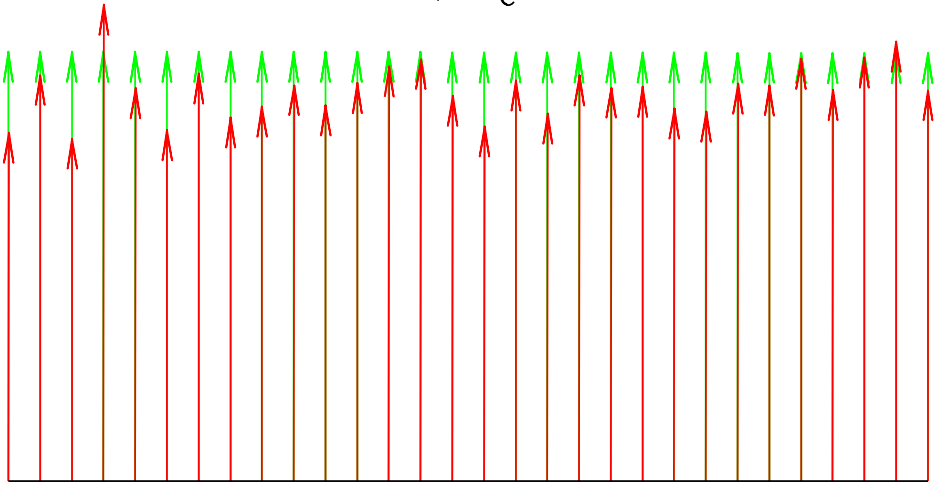
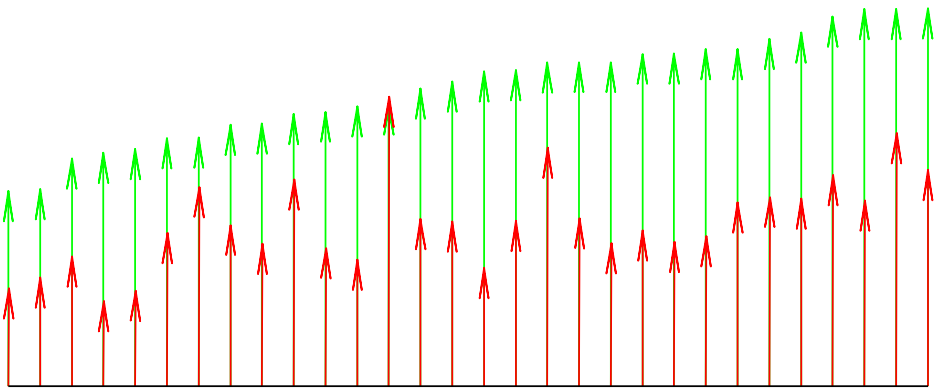
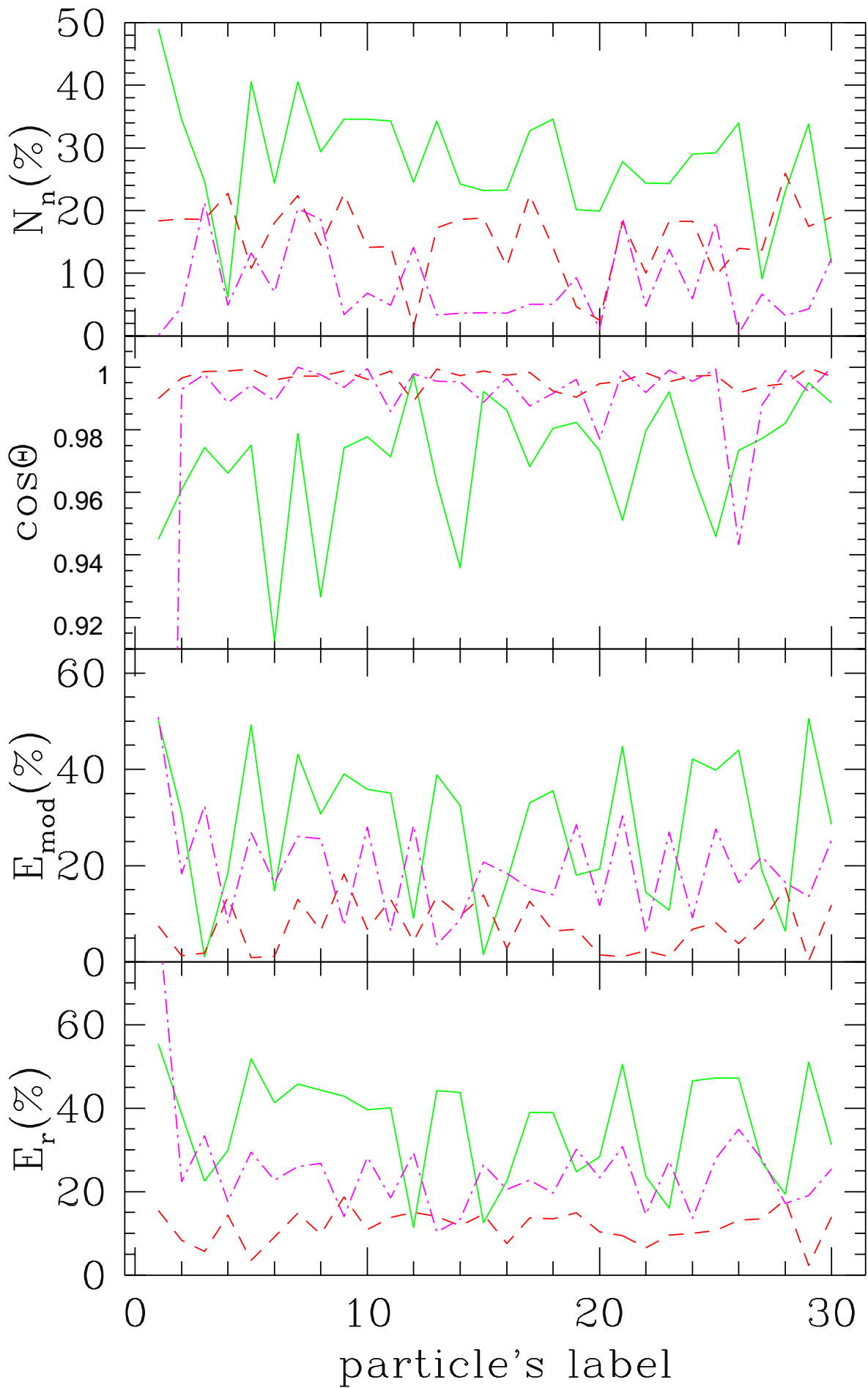
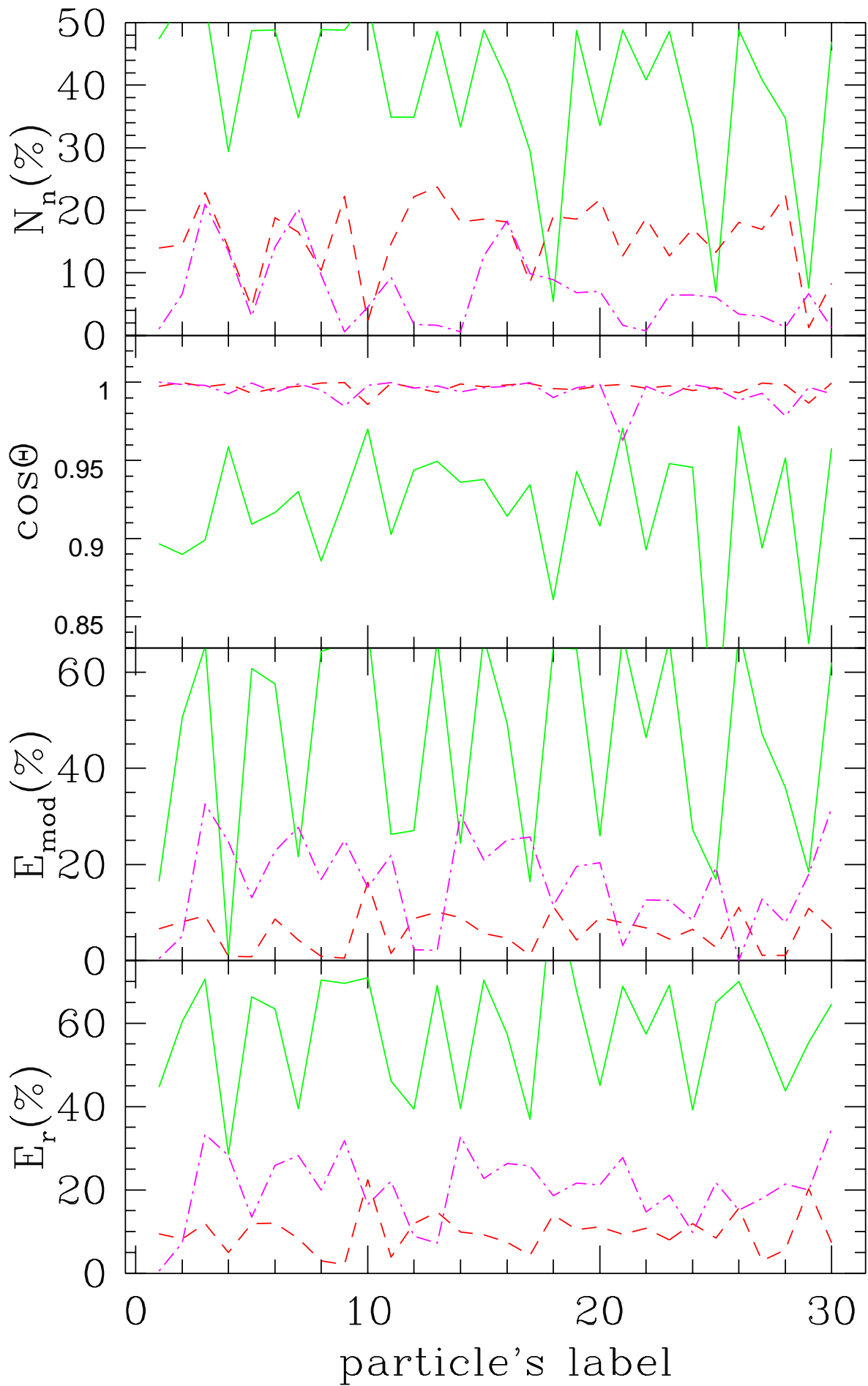
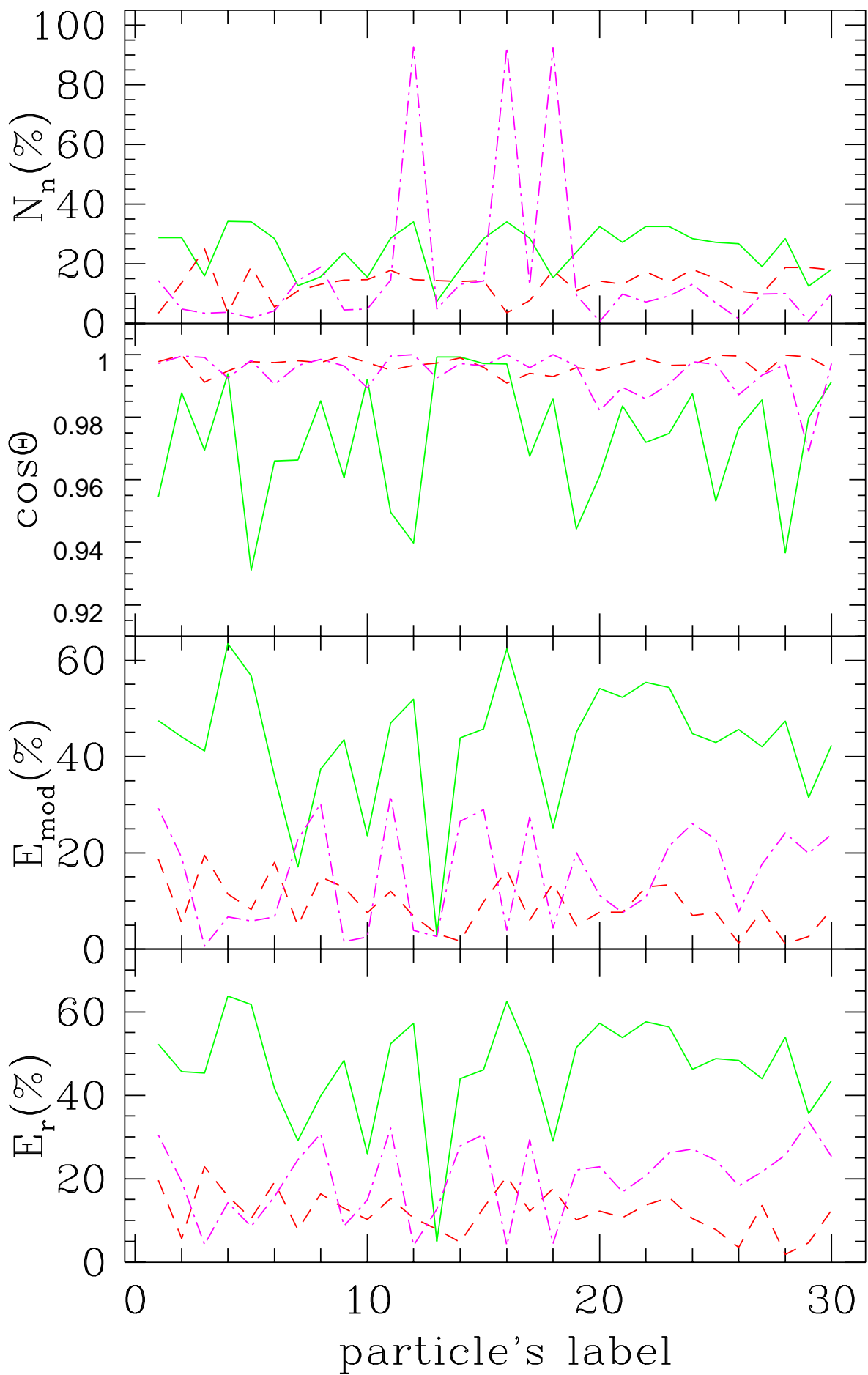
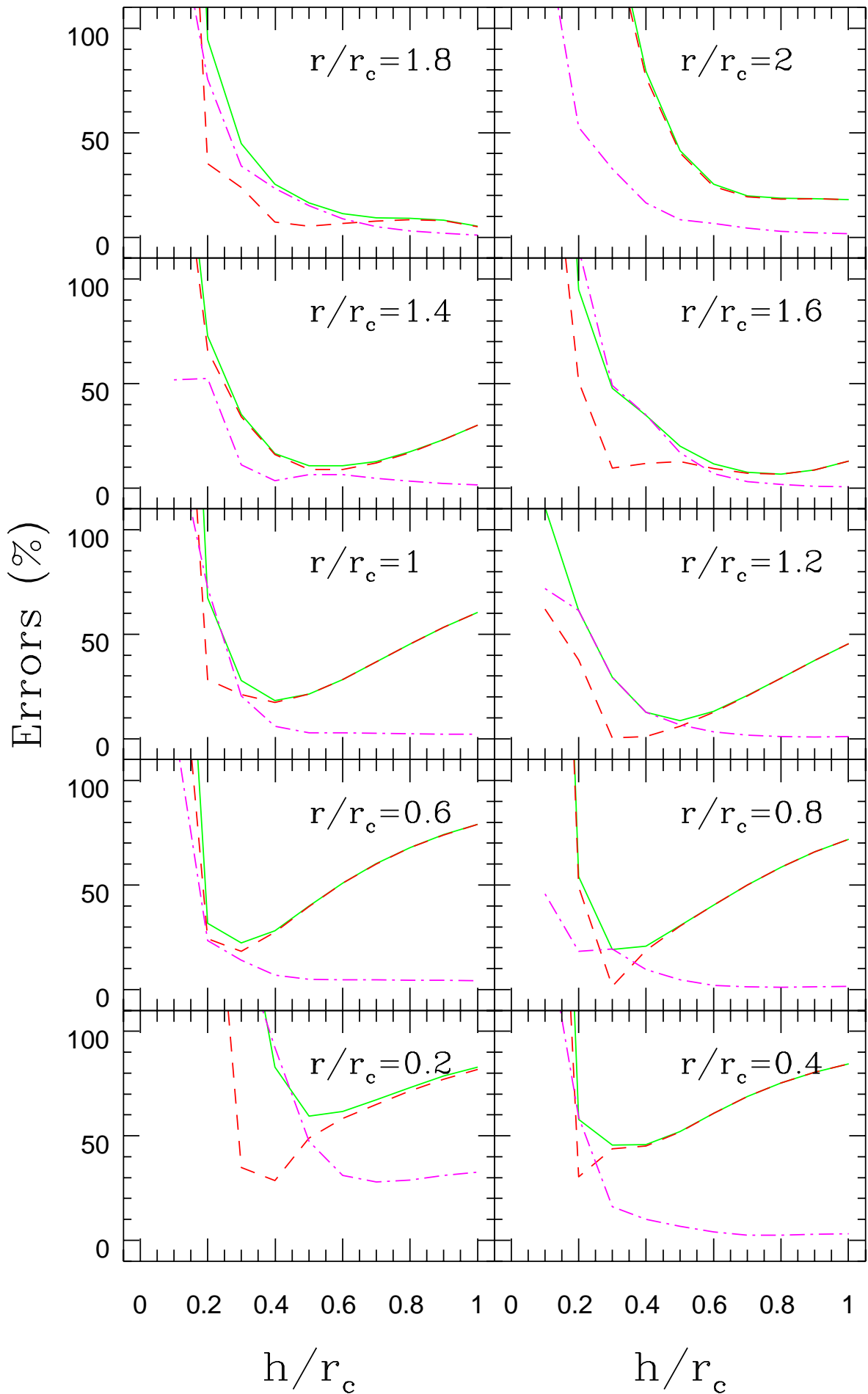
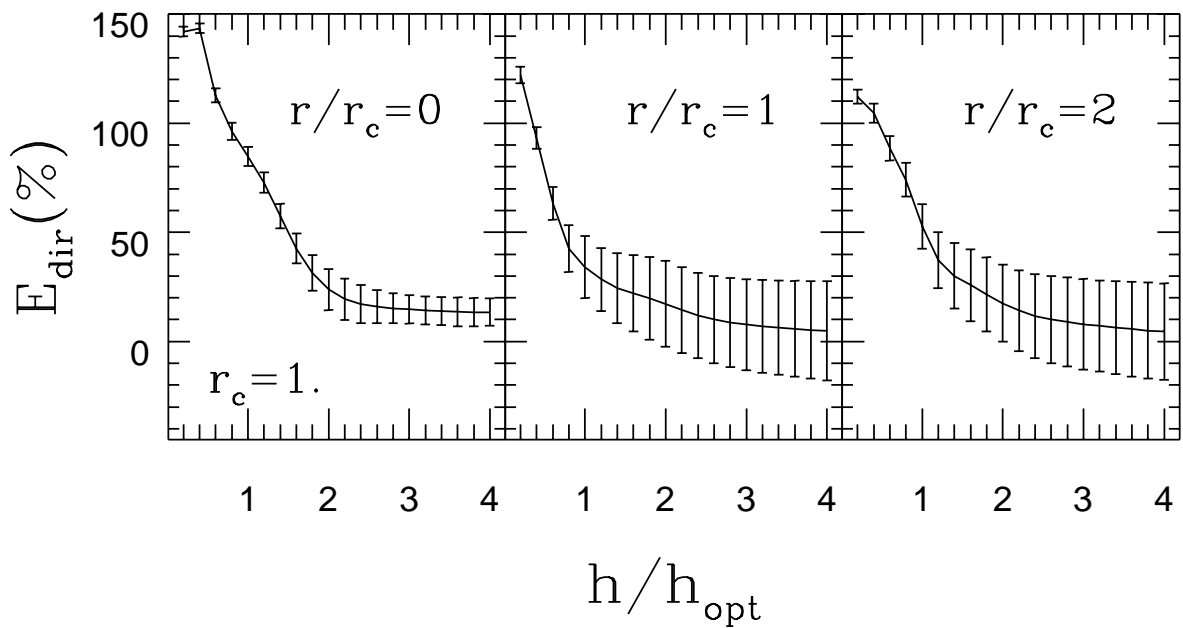
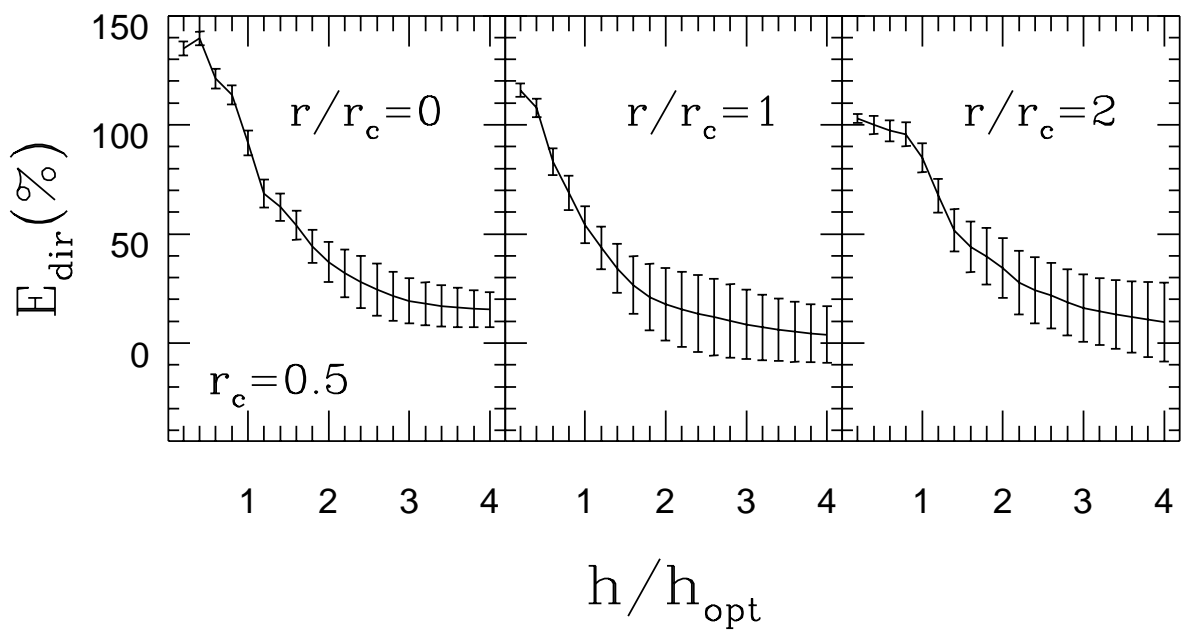
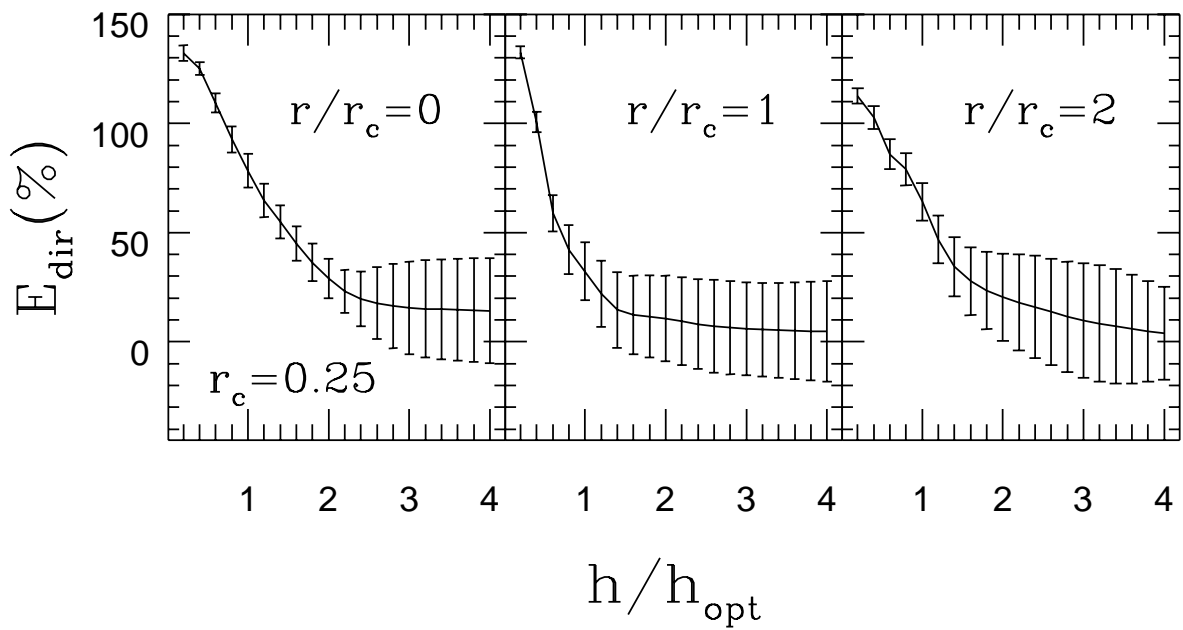
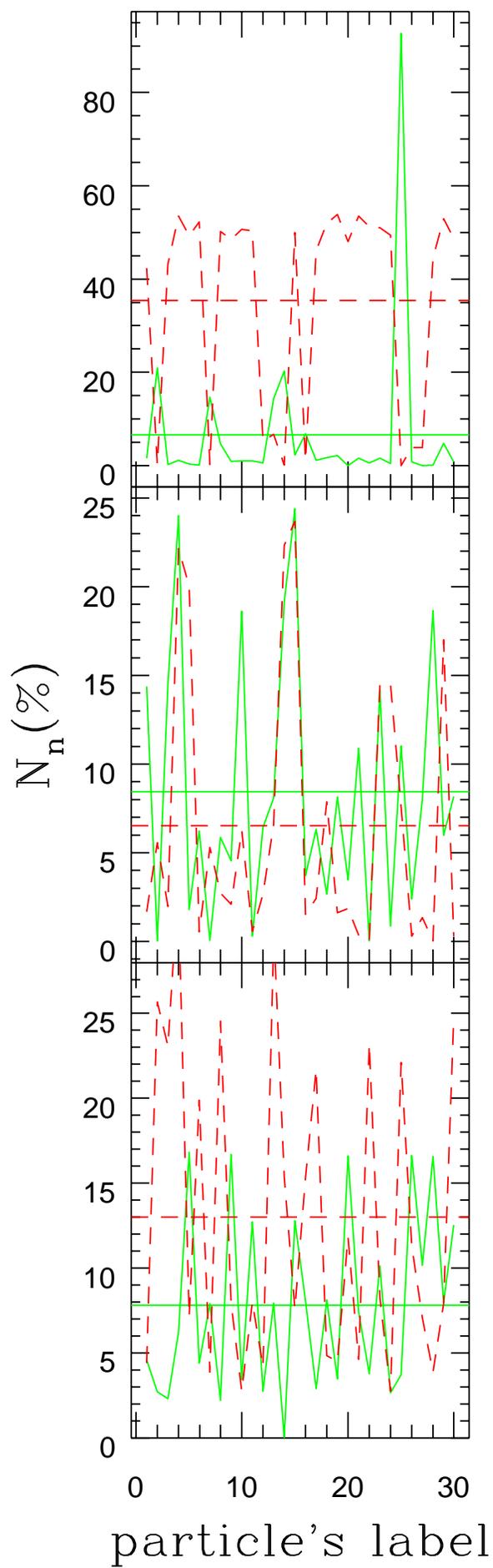
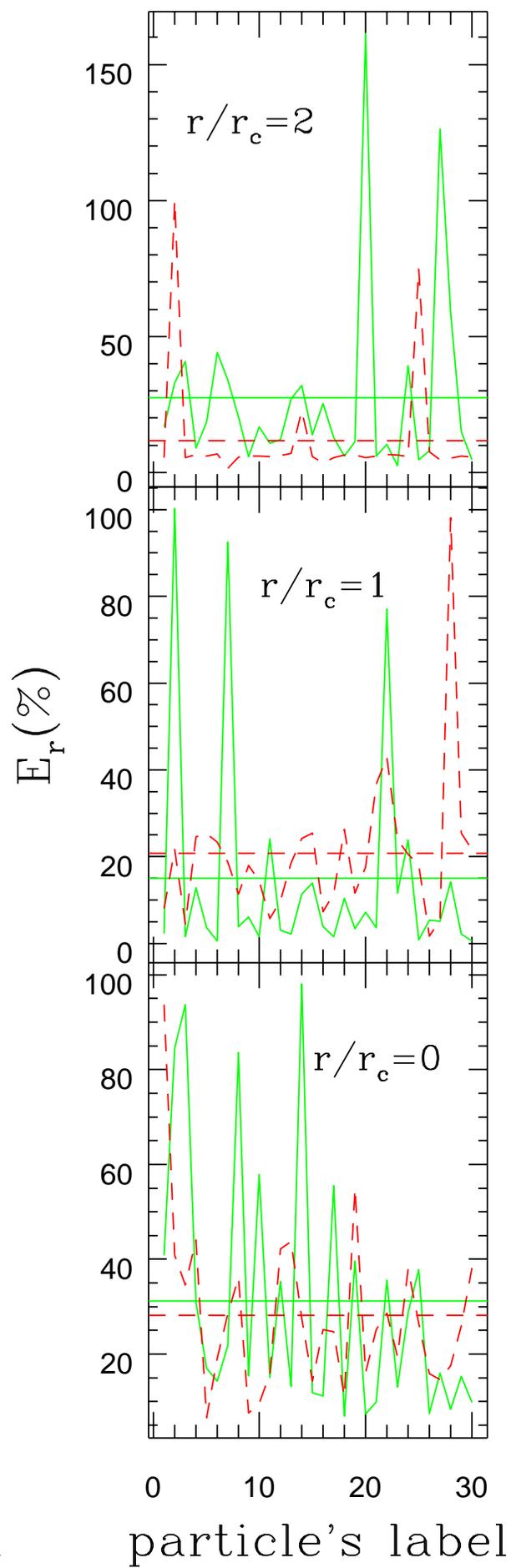
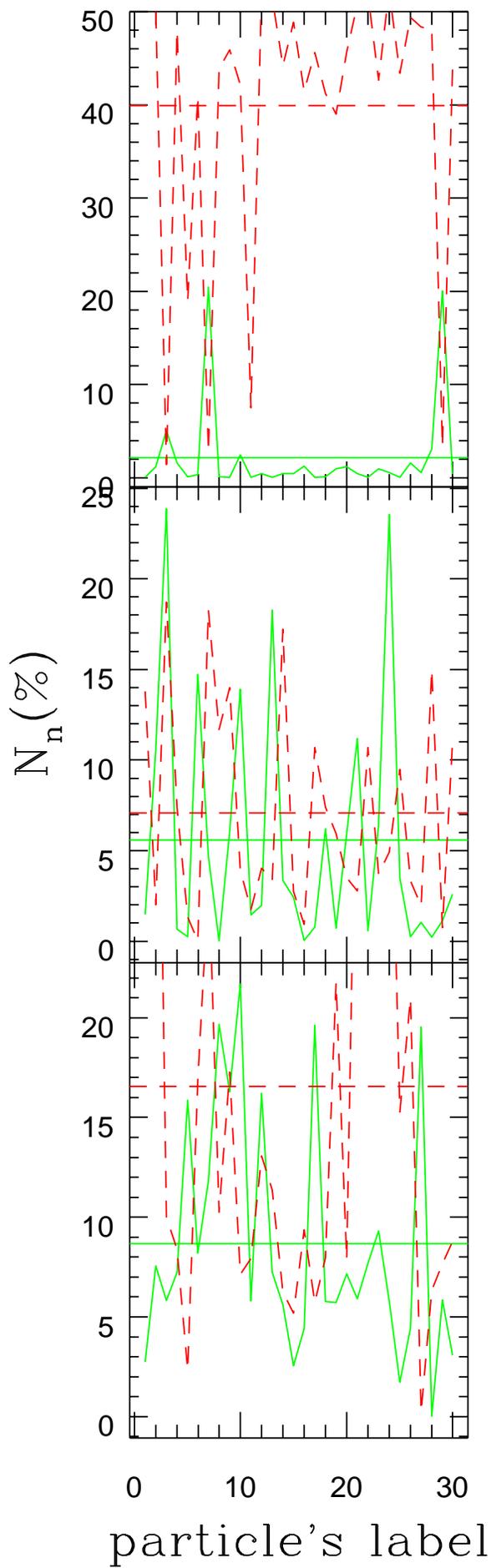
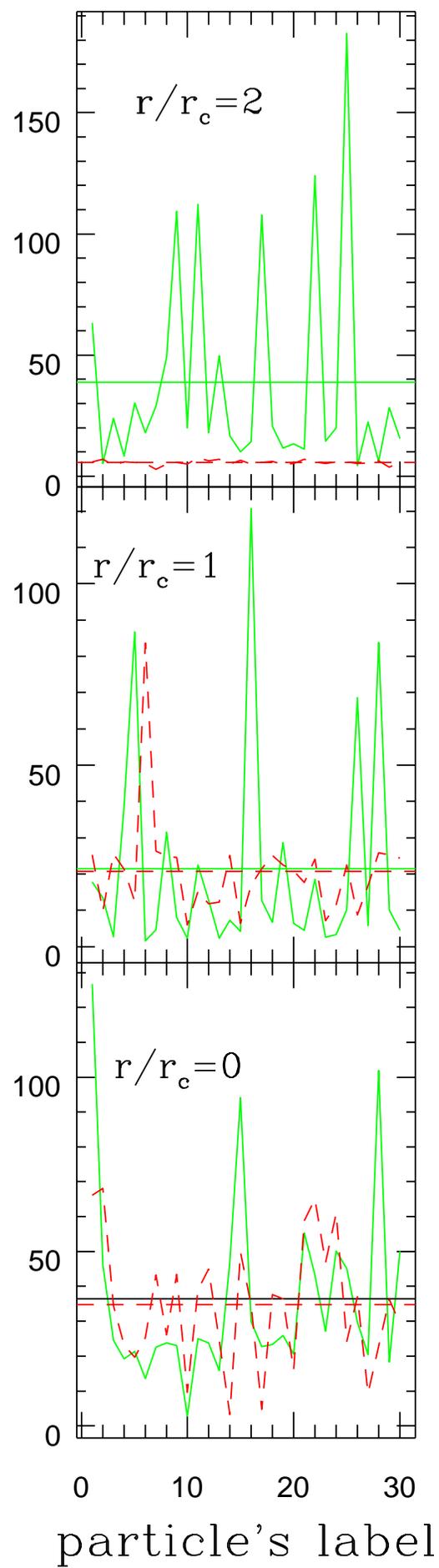
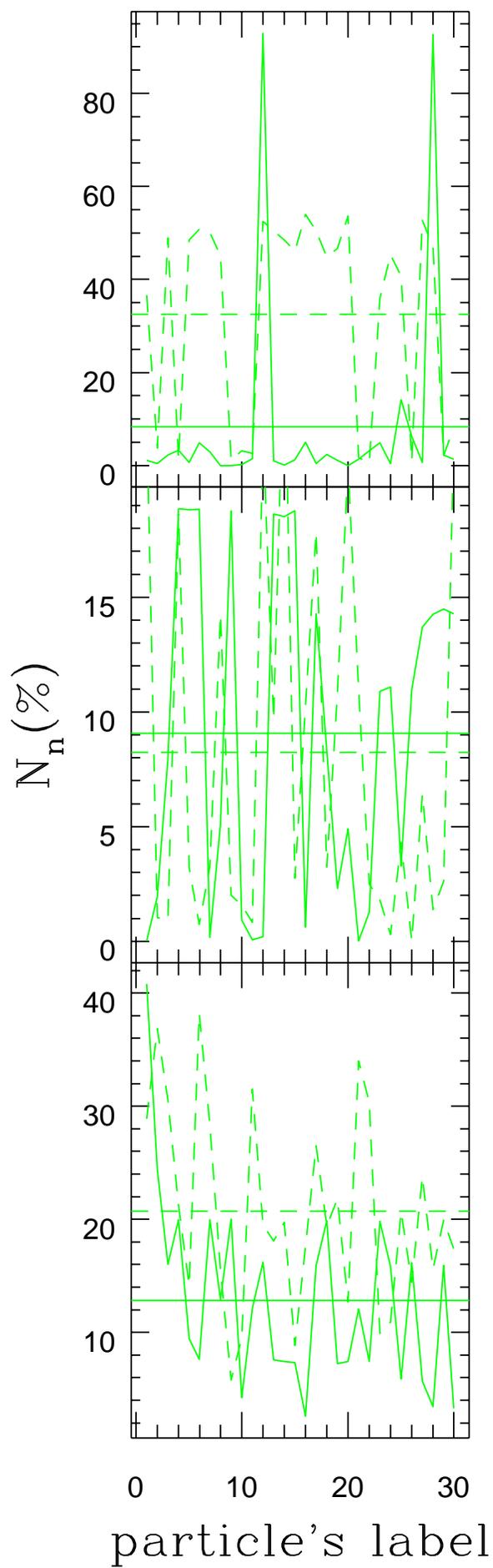
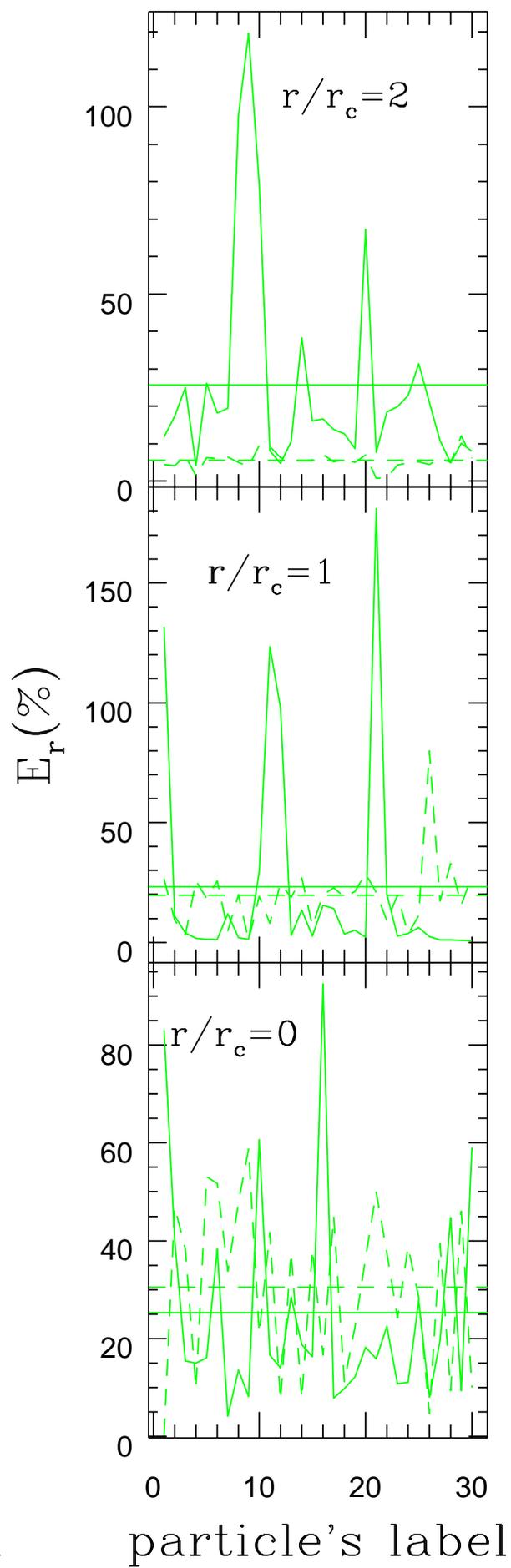
\end{document}